\documentstyle[12pt,epsfig,axodraw]{article}
\begin{document}
\newcommand{\newc}{\newcommand}
\newc{\beq}{\begin{equation}}
\newc{\eeq}{\end{equation}}
\newc{\barr}{\begin{eqnarray}}
\newc{\earr}{\end{eqnarray}}
\newc{\ra}{\rightarrow}
\newc{\lam}{\lambda}
\newc{\Lam}{\Lambda}
\newc{\eps}{\epsilon}
\newc{\gev}{\,GeV}
\newc{\tev}{\,TeV}
\newc{\Hbar}{{\bar H}}
\newc{\Ubar}{{\bar U}}
\newc{\Dbar}{{\bar D}}
\newc{\Ebar}{{\bar E}}
\newc{\eg}{{\it e.g.}\ }
\newc{\ie}{{\it i.e.}\ }
\newc{\nonum}{\nonumber}
\newc{\lab}[1]{\label{eq:#1}}
\newc{\labf}[1]{\label{fig:#1}}
\newc{\eq}[1]{(\ref{eq:#1})}
\newc{\eqs}[2]{(\ref{eq:#1},\ref{eq:#2})}
\newc{\fig}[1]{\ref{fig:#1}}
\newc{\rpv}{{\not \!\! R_p}}
\newc{\rpvm}{{\not \! R_p}}
\newc{\rp}{R_p}
\newc{\gsim}{\stackrel{>}{\sim}}
\newc{\lsim}{\stackrel{<}{\sim}}
\newc{\breq}{{\cal B}_{eq}}
\newc{\chargino}{{\tilde \chi}^{+}}
\newc{\charginol}{{\chi}^{+}_{l}}
\newc{\etal}{{\it et al.}\ }
\newc{\neutralino}{{\tilde \chi}^{0}}
\newc{\ms}{\mbox{$m_{\scriptstyle S}$}}
\newc{\half}{\frac{1}{2}}
\newc{\ol}{\overline}
\newc{\plb}[3]{Phys. Lett. B {#1} (19{#2}) #3}
\newc{\npb}[3]{Nucl. Phys. B {#1} (19{#2}) #3}
\newc{\prd}[3]{Phys. Rev D {#1} (19{#2}) #3}
\newc{\zpc}[3]{Z. Phys. C {#1} (19{#2}) #3}

\title{Hide and Seek with Supersymmetry} 
\author{Herbi Dreiner}
\date{\small{ Rutherford Laboratory, Chilton, Didcot, OX11 0QX, UK} }
\maketitle

\begin{abstract}
\noindent This is the summary of a 90 minute introductory talk on 
Supersymmetry presented at the August 1998 Zuoz Summer School on
``Hidden Symmetries and Higgs Phenomena''. I first review the
hierarchy problem, and then discuss why we expect supersymmetry just
around the corner, \ie at or below $1\tev$. I focus on the specific
example of the anomalous magnetic moment of the muon to show how
supersymmetry can indeed hide just around the corner without already
having been detected.  An essential part of supersymmetry's disguise
is the fact that it is broken. I end by briefly outlining how the
disguise itself is also hidden.
\end{abstract}
\vspace{2cm}

\section{Introduction}
In the Standard Model (SM), $SU(2)_L$ transformations relate the components
of multiplets
\beq
T^a:\quad e^-_L \longleftrightarrow \nu_e.
\eeq
Here $T^a$ is the $SU(2)_L$ isospin-one, bosonic generator, which
changes $SU(2)_L$ isospin. In supersymmetry the transformations relate
particles of different spin
\beq
Q_\alpha:\quad e^-_L(s=\half) \longleftrightarrow {\tilde e}^-_L(s=0),
\eeq
in this case the left-handed electron and the left-handed scalar electron
(selectron).\footnote{The selectron is a scalar and thus has no chirality.  
But in the SM chirality has become associated with $SU(2)_L$ isospin
and thus the partner of the $SU(2)_L$ doublet electron is called the
{\it left-handed} selectron.}  Here $Q_\alpha$ is a spinorial generator, a
2-component, spin-$1/2$, Weyl-spinor, which transforms as a left-handed spinor
under the Lorentz group. $\alpha=1,2$ is the spinor index.  $Q_\alpha$ changes
the spin and hence it must have a non-trivial algebra with the Lorentz-group. 
\barr
[P^\mu,Q_\alpha] &=&0 \lab{alg1}, \\
\left[M^{\mu\nu},Q_\alpha\right] &=&-i (\sigma^{\mu\nu})_
\alpha^\beta Q_\beta, \lab{alg2}\\
\{Q_\alpha ,Q_\beta\}=\{{\ol Q}_{\dot\alpha} ,
{\ol Q}_{\dot\beta}\}&=&0, \lab{alg3}\\
\{Q_\alpha,{\ol Q}_{\dot\beta}\}&=& 2 \sigma_{\alpha\dot\beta}^\mu
P_\mu. \lab{alg4}
\earr
Here $\sigma^\mu=(\bf{1},\sigma^i)$ and $\sigma^i$ are the Pauli
matrices. ${\ol\sigma}^\mu=(\bf{1},-\sigma^i)$ and $\sigma^{\mu\nu}=
\frac{1}{4}({\sigma}^\mu{\ol\sigma}^\nu-{\sigma}^\nu{\ol\sigma}^\mu)$.
$P^\mu$ is the energy-momentum tensor and $M^{\mu\nu}$ is the angular
momentum tensor. Since $Q_\alpha$ has spin and thus angular momentum
the commutator with $M^{\mu\nu}$ is non-trivial. The dotted spinorial
indices, \eg $\dot\beta$, transform as a right-handed spinors under the
Lorentz group.

From the algebra we can immediately derive some simple consequences.
\begin{itemize}
\item From Eq.\eq{alg3} and Eq.\eq{alg4} it follows that $Q_\alpha$ is a
raising operator and ${\ol Q}_{\dot\beta}$ is a lowering operator.
Since $Q_\alpha Q_\alpha=0$ (no summation) by Eq.\eq{alg3} we only have
two components to a supermultiplet\footnote{For an explicit proof
see Section 1.4 in \cite{bailin}.}
\beq
\{{\tilde e}^-_L\;(s=0);e_L^-\;(s=\half)\},\quad
\{{\tilde e}^-_R\;(s=0);e_R^-\;(s=\half)\},\quad
\{\lam_{\tilde \gamma}\;(s=\half);\gamma \;(s=1)\}.
\eeq
And each multiplet has equal numbers of bosonic and fermionic degrees
of freedom.
\item From Eq.\eq{alg1} we have $[P\cdot P,Q_\alpha]=0$ and thus
$[M^2,Q_\alpha]=0$, where $M$ is the mass of the field.  If
supersymmetry is conserved
\beq 
M({\tilde e}^-)=M(e^-)=511\,KeV,\quad M({\tilde \gamma}^-)=M(\gamma)=0.  
\eeq 
No charged scalar has been observed up to LEP2 energies, which are
significantly higher than $M(e^-)$. We therefore conclude that
supersymmetry must be broken.
\end{itemize}
These two points can be summarized into one formula \cite{ferrara}
\barr
STr {\cal M}^2&=&0,\\
\lab{supertr}
STr({\cal M}^2)&\equiv&\sum_i (-1)^{2S_i}\cdot (2S_i+1)\,  M_i^2.
\earr
$STr({\cal M}^2)$ is the supertrace of the mass matrix squared
containing fields of different spin, $S_i$. For just the electron
supermultiplet, we have
\beq
STr({\cal M}^2) = M_{{\tilde e}_L}^2 + M_{{\tilde e}_R}^2 -2 M_e^2=0, \qquad
\Longrightarrow\quad \frac{M_{{\tilde e}_L}^2 + M_{{\tilde e}_R}^2}{2}=M_e^2
\lab{electron}
\eeq
This is a catastrophe. At least one of the selectron fields must have
mass less than or equal to $M_e$, which is excluded by experiment. It
turns out that, Eq.\eq{supertr} is also true if global supersymmetry is
spontaneously broken, as we show in Sect.\ref{sect:susybr}. Therefore,
just breaking supersymmetry does not solve this problem of the
spectrum.  However, the supertrace formula need only apply to the full
set of fields. Thus, one might think a heavy selectron can be
compensated by a higgsino or gaugino which has opposite sign.  It has
proven impossible to construct such models and we shall come to this
problem again in Sect.\ref{sect:susybr}.

\section{Motivation for Supersymmetry}
To date there is no experimental evidence for supersymmetry. All the
same it is an intensely studied subject. Why? We first give an aesthetic 
argument, which however tells you nothing about the scale of supersymmetry
breaking. Next we discuss the hierarchy problem which indicates that
supersymmetry is broken near the electroweak scale.

\subsection{Unique Extension of the Symmetries of the S-Matrix}
The S-matrix in the Standard Model has {\it external} and {\it
internal} symmetries. The Lorentz invariance is a space-time symmetry
and thus an external symmetry.  The gauge symmetry $G_{SM}=SU(3)\times
SU(2)\times U(1)$ and the global baryon- and lepton-number invariance
are internal symmetries whose generators commute with the Lorentz
generators. The following remarkable theorem holds
\cite{coleman,haag}: Supersymmetry is the only possible external
symmetry of the S-matrix, {\it beyond} the Lorentz symmetry, for which
the S-matrix is not trivial. It is beyond the scope of this lecture to
present a formal proof. Instead, I give a heuristic argument which
hopefully makes the theorem at least plausible \cite{rossbook}.
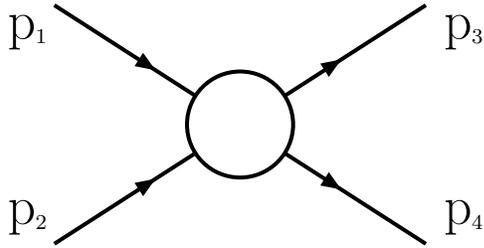
\begin{figure}[t]
\begin{center}
\begin{picture} (225,142)(0,0)
%\thicklines
\SetWidth{1.5}
\SetScale{1.}
% incoming fermions
\ArrowLine(5,5)(75,50)
\ArrowLine(5,95)(75,50)
%outgoing particles 
\ArrowLine(75,50)(145,95)
\ArrowLine(75,50)(145,5)
% circle around scattering point
\BCirc(75,50){20}
\Text(-5,85)[]{{\LARGE p}$_1$}
\Text(-5,15)[]{{\LARGE p}$_2$}
\Text(160,85)[]{{\LARGE p}$_3$}
\Text(160,15)[]{{\LARGE p}$_4$}
\end{picture}
\end{center}
\caption{{\it Spinless $2\ra2$ scattering. The $p_i$ label the momenta of
the particles.}}
\labf{spinless}
\end{figure}

Consider spin-less $2\ra2$ scattering as shown in Fig. \fig{spinless}.
The incoming four-momenta are labelled by $p_1,\,p_2$ and the outgoing
four-momenta by $p_3,\,p_4$. Momentum conservation (Lorentz symmetry)
implies $p_1+p_2=p_3+p_4$. For simplicity, we shall assume the
particles have equal mass $p_i^2=m_i^2=m^2$. The argument below
also holds for more general masses. The scattering amplitude is a
Lorentz scalar. It can thus only depend on the Lorentz invariants of
the scattering process. The most common choice of these invariants are
the Mandelstam variables
\beq
s=(p_1+p_2)^2,\quad t=(p_1-p_3)^2,\quad u= (p_2-p_3)^2.
\eeq
Since $s+t+u=4m^2$, only two of the invariants are independent. 
We choose the centre-of-mass energy squared, $s$, and the scattering 
angle $\cos\theta=1-2t/(s-4m^2)$ in the centre-of-mass system.

According to the Coleman-Mandula theorem \cite{coleman}, if we add any
further {\it external} symmetry to our theory, whose generators are
tensors, then the scattering process must be trivial, \ie there is no
scattering at all. Consider the following example. Assume the tensor
$R^{\mu\nu} = p^\mu p^\nu-\frac{1}{4} g^{\mu\nu}m^2$, is conserved,
where $\mu,\nu=1,\ldots,4$ are Lorentz vector indices. Then
\barr
R_1^{\mu\nu} +R_2^{\mu\nu}&=& R_3^{\mu\nu}+R_4^{\mu\nu},\\
\Longrightarrow\quad  
p_1^{\mu}p_1^{\nu} + p_2^{\mu}p_2^{\nu} &=& p_3^\mu p_3^\nu + 
p_4^\mu p_4^\nu. \lab{rconserv}
\earr
In the centre-of-mass coordinate system we have
\beq
\begin{array}{ll}
p_1=(E,0,0,p),& p_2=(E,0,0,-p), \\ 
p_3=(E,0,p\sin\theta,p\cos\theta),&
p_4=(E,0,-p\sin\theta,-p\cos\theta).
\end{array}
\eeq
The conservation of $R^{\mu\nu}$ \eq{rconserv} then implies for
$\mu=\nu=4$ 
\beq 
2p^2=2p^2\cos^2\theta,\quad \Longrightarrow \quad \theta=0.
\eeq
There is thus no scattering at all if $R^{\mu\nu}$ is conserved. The
more general statement is that {it any} new conserved external tensor
leads to trivial scattering.

Coleman and Mandula \cite{coleman} showed that the only possible {\it
conserved} quantities that transform as {\it tensors} under the
Lorentz group are the generators of the Poincar{\'e}\ group
$P_\mu,\,M_{\mu\nu}$ and Lorentz scalars $C_i$. Tensors are
combinations of vector indices and are thus bosons. The argument of
Coleman-Mandula does {\it not} apply to conserved charges transforming
as {\it spinors}. This is just the case of supersymmetry.  Haag,
Lopuszinski, and Sohnius showed that when extending the Lorentz
algebra to include a single spinorial charge $Q_\alpha$ the algebra
\eq{alg1}-\eq{alg4} is unique \cite{haag}. Thus supersymmetry occupies
a very special place with respect to the Lorentz group, a fundamental
symmetry in nature. It is the {\it unique} extension of the external
(Lorentz) symmetry which still allows for a non-trivial S-matrix.  It is
thus tantalizing to enquire whether supersymmetry is realized in
nature. From the arguments above, we know that if it {\it is} realized, it
must be broken.  We must therefore first understand at what scale
supersymmetry might be broken.

\subsection{Hierarchy Problem}
In order to discuss the hierarchy problem \cite{gildener}, I focus on
the relevant two-point functions for vanishing external
momenta\footnote{In this subsection I have relied heavily on the very
nice discussion by Manuel Drees \cite{drees}. In just one longer
lecture I can only give a flavour of the problem. For more detail
please consult his lecture notes \cite{drees}.}.  Consider the
two-point function of the photon at one-loop for vanishing external
momenta as shown in Fig. \fig{2pt}a
\barr
\Pi_{\gamma\gamma}^{\mu\nu}(0)&=&-\int \frac{d^4k}{(2\pi)^4} {\rm
Tr}\left[ (-ie\gamma^\mu)\frac{i}{\not\! k -
m_e}(-ie\gamma^\nu)\frac{i}{\not \!k - m_e} \right],\nonumber \\ &=&0. 
\earr 
The vanishing of this correction is guaranteed by gauge invariance,
\ie the mass of the photon is protected by a symmetry. In order to get
the correct result, one must perform the calculation in a
regularization scheme which preserves gauge invariance such as
dimensional regularization. It is therefore not possible to study the
dependence as a function of a high-energy momentum cut-off.

\begin{figure}[t]
\begin{center}
\begin{picture} (300,45)(0,0)
%\thicklines
\SetWidth{1.}
\SetScale{1.3}
%
%Photon Self Energy
% 

\Photon(-40,15)(-10,15){2}{4}
\Photon(20,15)(50,15){2}{4}
\ArrowArc(5,15)(15,0,180)
\ArrowArc(5,15)(15,180,0)
\Vertex(-10,15){1.5}
\Vertex(20,15){1.5}
\Text(-45,30)[c]{$\gamma$}
\Text(55,30)[c]{$\gamma$}
\Text(5,47)[c]{$e$}
\Text(5,10)[c]{$e$}
%
% Electron Self-energy
% 
\ArrowLine(80,15)(108,15)
\ArrowLine(108,15)(141,15)
\ArrowLine(141,15)(174,15)
\Vertex(108,15){1.5}
\Vertex(141,15){1.5}
\PhotonArc(124.5,15)(16.5,0,180){2}{6.5}
\Text(105,27)[c]{$e$}
\Text(220,27)[c]{$e$}
\Text(164.5,50)[c]{$\gamma$}

% Scalar Self Energy
\DashLine(190,15)(220,15){3}
\DashLine(250,15)(280,15){3}
\ArrowArc(235,15)(15,0,180)
\ArrowArc(235,15)(15,180,0)
\Vertex(220,15){1.5}
\Vertex(250,15){1.5}
\Text(250,30)[c]{$\phi$}
\Text(360,30)[c]{$\phi$}
\Text(310,51)[c]{$f$}
\Text(310,10)[c]{$f$}
% Labels
\Text(5,-15)[]{{(a)}}
\Text(165,-15)[]{{(b)}}
\Text(310,-15)[]{{(c)}}
\end{picture}
\end{center}
\caption{{\it (a) The photon two point function, (b) The electron two
point function and (c) the scalar two point function.}}
\labf{2pt}
\end{figure}
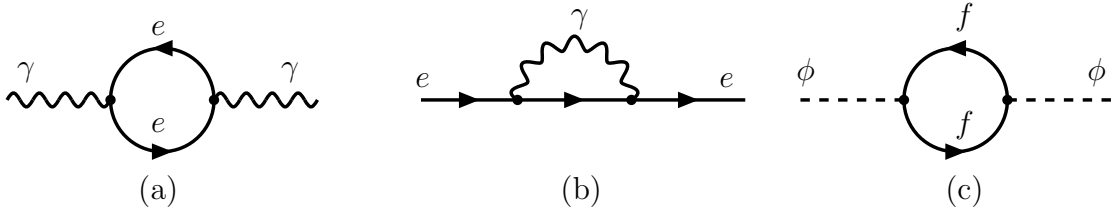
The diagram for the one-loop correction to the electron two-point
function is given in Fig. \fig{2pt}b. The calculation gives
\barr
\Pi_{ee}(0)&=&\int \frac{d^4k}{(2\pi)^4}(-ie\gamma^\mu)\frac{i}{\not \!k
- m_e}(-ie\gamma^\nu) \frac{-ig^{\mu\nu}}{k^2},\nonumber
\\ &=&-4e^2m_e \int
\frac{d^4k}{(2\pi)^4} \frac{1}{k^2(k^2-m_e^2)}. 
\lab{e2pt}
\earr 
If we consider the SM as an effective theory which is valid to a high
energy scale $\Lam$, then we must cut-off the integral at $\Lam$.  The
expression in Eq.\eq{e2pt} diverges logarithmically as the cutoff $\Lam$
of the integral goes to infinity. Evaluating the integral for a fixed
value of $\Lam$, the 1-loop correction to the mass of the electron is
given by
\beq 
\Delta m_e\approx
\frac{2\alpha_{em}}{\pi} m_e \log\frac{\Lam}{m_e} \approx 0.24\;m_e,
\eeq 
In the last step we have inserted the maximal value $\Lam=M_{Pl}$,
where $M_{Pl}={\cal O}(10^{19}\gev)$ is the Planck scale, where
gravitational effects become strong.
The mass correction is thus moderate even in the extreme case of
requiring the SM to be valid to the Planck scale. It is important also
to notice that the correction vanishes for vanishing electron
mass. Thus a vanishing fermion mass is stable under quantum
corrections. This is because for $m_e=0$ the Lagrangian is invariant
under the additional chiral transformations $\psi_e\ra\psi_e'=e^{i\alpha
\gamma_5}\psi_e$, \ie the fermion mass is protected by a symmetery.

For a scalar field the 1-loop Feynman graph is shown in
Fig. \fig{2pt}c.  and the two-point function is given by
\beq
\Pi_{\phi\phi}^f(0)=-2N(f)\lam_f^2 \int \frac{d^4k}{(2\pi)^4}
\left[ \frac{1}{k^2-m_f^2}+\frac{2m_f^2}{(k^2-m_f^2)} \right].
\lab{smphi}
\eeq
Here $N(f)$ is the multiplicity of the fermion coupling to the scalar,
\eg the colour degeneracy.  $\lam_f$ is the Yukawa coupling for the
interaction $\phi{\ol\psi}_f\psi_f$. The second term in the integrand
gives a logarithmically divergent part as well as a finite
correction. The first part of the integrand diverges quadratically
with the momentum cut-off $\Lam$. If we insert $\Lam=M_{Pl}$ then the
mass correction to the scalar field is of order the Planck scale
\beq
\Delta m^2_\phi \approx (10^{19}\, GeV)^2.
\eeq
However, the SM requires $m_\phi \lsim {\cal O}(1\tev)$. Thus this
correction is completely unacceptable. This is called the hierarchy
problem because the extreme difference (hierarchy) in energy scales in
the theory is inconsistent in the fundamental scalar sector. 

In addition to the divergent term there is a finite term in Eq.\eq{smphi}
\beq
(\Delta m^2_\phi)_{finite}=\frac{N(f)\lam_f^2}{8\pi}m_f^2.
\lab{finite}
\eeq
If there is a heavy fermion at the scale $\Lam$ with $m_f= {\cal
O}(\Lam)$ which couples to the scalar field $\phi$ (possibly via
loops) then according to Eq.\eq{finite} this gives a correction of order
$\Lam^2$, just as the divergent term. Both can be cancelled by a
counter-term $\delta m_\phi^2={\cal O}(\Lam^2)$.  However, the
coefficients of the counter term, of the finite term in Eq.\eq{finite} as
well as the $\Lam^2$ divergent term from the integral must cancel to
$M_W^2/\Lam^2$ or approximately one part in $10^{28}$. This is a huge
fine-tuning and is extremely unnatural. It must also be achieved in
each order of perturbation theory separately, \ie the coefficients
must be re-tuned in each order of perturbation theory. Furthermore, we
expect any more fundamental theory, \eg M-theory to predict some or
all low-energy parameters. Such a theory must predict the coefficients
of the scalar mass correction to one part in $10^{28}$ in order to
obtain the correct low-energy theory. It will always be an imposing
challenge to construct such a theory.
%%%%%%%%%%%%%%%%%%%%%%%%%%%%%%%%%%%%%%%%%%%%%%%%
\begin{figure}[h,t]
\begin{center}
\epsfig{file=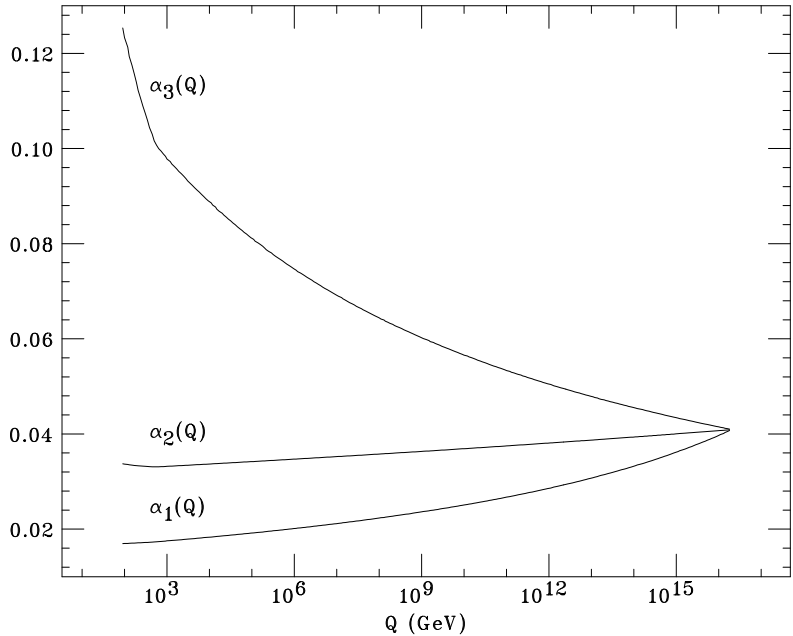,width=8.5cm,height=10.5cm,angle=90}
\end{center}
\vspace*{-5mm}
\caption[dummy]{{\it The evolution of the coupling constants in the
MSSM at two-loop with threshold corrections for the weak scale
supersymmetric fields.}}
\labf{unification}
\vspace*{-5mm}
\end{figure}
%%%%%%%%%%%%%%%%%%%%%%%%%%%%%%%%%%%%%%%%%%%%%%%%
In the formulation of the hierarchy problem it is essential that there
is a higher scale $\Lam$ (=$M_{GUT},\,M_{Pl},\,M_X$) where there is
new physics, \eg heavy fermions. If the SM is valid for $\Lam
\ra\infty$ then there is {\it no} hierarchy problem. One could for
example perform the previous calculations of the two-point functions
in dimensional regularization. The divergences are then renormalized
and no quadratic divergence appears anywhere in the calculation.  This
is perfectly consistent. So the question arises: is there a new scale
of physics? The answer is most likely, yes. There are several hints
of a new higher scale of physics

\begin{itemize}
\item Newton's constant
indicates that there is a scale $M_{Pl}$ where the quantum effects of
gravity are significant. At this scale we expect the physics content
to change and thus our SM calculation to break down. 
\item The observed coupling constant unification in supersymmetry, as
seen in\footnote{I am grateful to Athansios Dedes for providing me
with this up-to-date figure.}  Fig. \fig{unification} strongly hints
towards two new scales of physics: the supersymmetry scale and the
unification scale.
\item The claimed observation of neutrino mass by the Super Kamiokande
experiment \cite{superk} is most easily explained via the see-saw
mechanism which requires a new high scale of physics.
\item In the SM, electroweak baryogenesis does not supply a large enough baryon
asymmetry \cite{gavela}. Thus there must be a new scale of physics where
baryon-number is violated {\it and} a sufficient baryon-asymmetry is
generated. It is worth pointing out that in supersymmetry baryogenesis
at the electroweak phase transition is still possible \cite{carlos}.
\end{itemize}

%%%%%%%%%%%%%%%%%%%%%%%%%%%%%%%%%%%%%%%%%%%%%%%%%
\begin{figure}[h,t]
\begin{center}
\epsfig{file=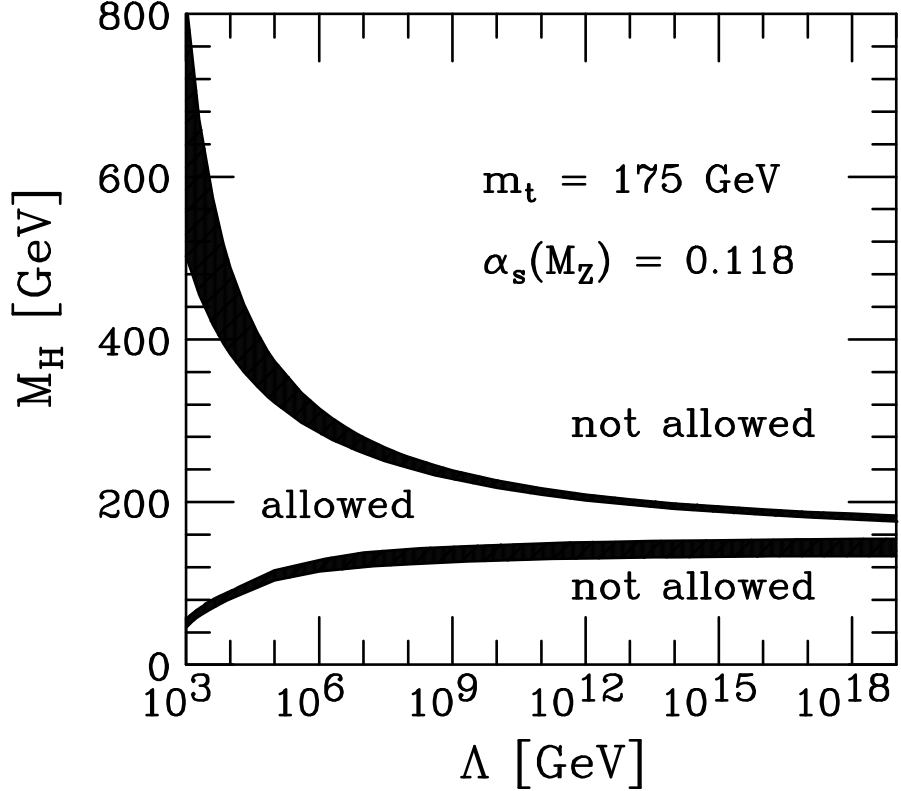,width=10.5cm,angle=90}
\end{center}
\vspace*{-5mm}
\caption[dummy]{{\it Theoretical bounds on the Higgs mass when considering
the SM as an effective theory valid to a cut-off scale $\Lam$
\cite{riesselmann}. The thickness of the curves is due to theoretical 
uncertainties.}}
\labf{higgsb}
\vspace*{-5mm}
\end{figure}
%%%%%%%%%%%%%%%%%%%%%%%%%%%%%%%%%%%%%%%%%%%%%%%%%

The idea of a new scale can be experimentally tested in the Higgs
sector. Within the SM only a certain range of Higgs masses are allowed
given a cutoff $\Lam$. An upper bound is obtained from the requirement
that the Higgs interactions are perturbative below $\Lam$. A lower
bound is obtained requiring the SM minimum of the Higgs potential to
be the true minimum for scales below $\Lam$. These bounds are shown
in\footnote{I thank Kurt Riesselmann for kindly providing this
figure.}  Fig. \fig{higgsb} \cite{riesselmann}. Thus for example if a
SM Higgs boson is discovered at the LHC with a mass larger than
$400\gev$ there must be new physics below $\Lam=10^5\gev$. If the SM
is valid to the Planck scale, then the Higgs boson mass must lie in
the narrow range: $135-180\gev$. In this interesting case, it is most
likely found via the decay channel $H\ra W^+W^-\ra\ell^+\ell^-\nu\nu$
\cite{dittmar}.

\subsection{The Supersymmetric Solution to the Hierarchy Problem}
As we saw in the introduction, in supersymmetry every field has a
superpartner differing by half a unit of spin but otherwise with
identical quantum numbers. ($Q_\alpha$ commutes with the internal
gauge symmetries.)  Thus the field $\phi$ also couples to the
superpartners, ${\tilde f}_L$, ${\tilde f}_R$, of the fermion field $f$
giving new contributions to the two point function of $\phi$. The
$\phi{\tilde f}_i{\tilde f}_j$ interaction Lagrangian is given by
\barr
{\cal L}_{\phi{\tilde f}}=-\frac{{\lam}^2_f}{2}\phi^2(|{\tilde
f}_L|^2+ |{\tilde f}_R|^2)-v{\lam}^2_f \phi (|{\tilde
f}_L|^2+ |{\tilde f}_R|^2) + \left( \frac{\lam_f}{\sqrt{2}}\right)
A_f\phi{\tilde f}_L{\tilde f}_R + h.c.
\lab{intlag}
\earr
The coupling constant is the same as for $\phi ff$ in the SM due to
supersymmetry.  $v$ is the vacuum expectation value of the Higgs.  I
have included the last term which explicitly breaks supersymmetry. For
the two point function we obtain
\barr
\Pi_{\phi\phi}^{\tilde f}(0)&=&
\lam^2 N(f) \int \frac{d^4k}{(2\pi)^4}
\left[\frac{1}{k^2-m^2_{{\tilde f}_L}} + \frac{1}{k^2-m^2_{{\tilde f}_R}}
\right] 
\nonumber \\
&&+ (\lam^2_fv)^2 N(f)
\int \frac{d^4k}{(2\pi)^4}
\left[\frac{1}{(k^2-m^2_{{\tilde f}_L})^2} 
+ \frac{1}{(k^2-m^2_{{\tilde f}_R})^2}
\right]\nonumber \\
&&+ (\lam_fA_f)^2 )N(f) \int \frac{d^4k}{(2\pi)^4}
\frac{1}{(k^2-m^2_{{\tilde f}_L})(k^2-m^2_{{\tilde f}_R}) }.
\lab{susyphi}
\earr
The two terms in the first line exactly cancel the previous $\Lam^2$
divergence in Eq.\eq{smphi}, since by supersymmetry we have the identical
coupling $\lam_f$ and degeneracy $N({\tilde f}_L) = N({\tilde f}_L) =
N(f)$. If we combine Eq.\eq{smphi} and Eq.\eq{susyphi} we obtain \cite{drees}
\barr
\Pi_{\phi\phi}^{f+{\tilde f}}(0)&=&
i\frac{\lam_f^2N(f)}{16\pi^2}\left[
-2m_f^2\left(1-\log\frac{m_f^2}{\mu^2}\right)+4m_f^2\log\frac{m_f^2}
{\mu^2}\right. \nonumber \\
&&\left.+2m^2_{\tilde f}\left(1-\log\frac{m_{\tilde f}^2}{\mu^2}\right)
-4m_f^2\log\frac{m_{\tilde f}^2}{\mu^2}-|A_f|^2\log\frac{m_{\tilde 
f}}{\mu^2} \right],
\lab{sumcorr}
\earr
and the $\Lam^2$ divergent terms have indeed cancelled. Here $\mu$ is
the renormalization scale. In the exact supersymmetric limit $m_f=m_
{\tilde f}$ and $A_f=0$ and the total radiative correction vanishes.
If we allow for supersymmetry breaking, we see that the cancellation of
the quadratic divergence is independent of the value of $m_{\tilde f}$
and $A_f$.

\section{The MSSM}
The minimal supersymmetric standard model (MSSM) is the extension of
the SM to include supersymmetry. The particle content is given by
\vspace{-0.5cm}
\begin{table}[h]
\begin{center}
\begin{tabular}{|ccccc|cccc|cc|}
\hline
$(e,\nu_e)_L$ & $e^c_L$ &$(u,d)_L$ &$u^c_L$&$d^c_L$ 
&$\gamma$& $W^\pm$ &$Z^0$ &$g_{i=1,\ldots,8}$&$H_1$& $H_2$
\\ \hline
$({\tilde e},{\tilde\nu}_{e})_L$ &${\tilde e}^c_L$ &$
({\tilde u},{\tilde d})_L$ & ${\tilde u}^c_L$&${\tilde d}^c_L$  
&${\tilde\gamma}$& ${\tilde W}^\pm$ &${\tilde Z}^0$ &
${\tilde g}_{i=1,\ldots,8}$ &${\tilde H}_1$& ${\tilde H}_2$ \\
\hline
\end{tabular}
\end{center}
\end{table}

\vspace{-0.5cm}

{\noindent} The first row contains the fields of the SM: one family of
matter fields, the gauge fields and {\it two} Higgs doublets. An extra
Higgs doublet is required to cancel the mixed gauge anomalies from the
spin-$\half$ superpartners of the Higgs: the Higgsinos ${\tilde H}_1,
\;{\tilde H}_2$. In the second line we show the superpartners of the
above fields which differ by one-half a unit of spin. The partners of
the matter fields are scalars, and the partners of the gauge and Higgs
fields are spin one-half fermions. The supersymmetric Lagrangian for
this field content contains no new parameters beyond the SM except
$\tan\beta=v_2/v_1$, the ratio of the vacuum expectation value of the
two neutral Higgs fields. We assume here that R-parity is conserved. 
Otherwise there are additional Yukawa couplings \cite{dreiner} and 
the phenomenology dramatically changes \cite{rosshadron}. However, 
we know that supersymmetry is broken, and we therefore need 
additional mass terms.
\beq
\begin{array}{lll}
{\rm Scalars:}&  {\tilde m}^2 \phi^\dagger_{\tilde f} \phi_{\tilde f},
&{\tilde f}\;\epsilon\; ({\tilde e}_L,{\tilde e}_R,\ldots,{\tilde
t}_L,{\tilde t}_R), \\
{\rm  Gauginos:}   &     m\lam_{\tilde g}\lam_{\tilde G}, &
\lam_{\tilde G}\;\epsilon \; ({\tilde\gamma},{\tilde
W}^\pm, {\tilde Z}^0,{\tilde g}),\\
{\rm Higgsinos:} & m{\ol\psi}_H\psi_H, & H\;\epsilon\;({\tilde H}_1,
{\tilde H}_2).
\end{array}
\eeq
Experimentally, no superpartner has yet been observed and so there are
lower bounds on these mass terms. For example from LEP we
know\footnote{For the bottom squark we have used the bound for zero
mixing.} \cite{lepbound}
\beq
{\tilde m}_{e_L}\gsim 95\gev,\quad {\tilde m}_b\gsim 90\gev,
\quad {\tilde M}_{\chi^\pm_1}\gsim95\gev.
\eeq
As we saw in Eq.\eq{intlag}, besides the mass terms there are further
supersymmetry breaking terms: tri-linear and bi-linear scalar
couplings $A_{ijk}\phi_i\phi_j\phi_k$ and $B_{ij}\phi_i\phi_j$,
respectively. In Eq.\eq{sumcorr}, we saw that none of these terms affect
the cancellation of the quadratic divergences. There is thus no reason
to exclude them. In total there are 124 undetermined parameters in the
MSSM including 19 from the SM \cite{haber}. Many of these parameters
are very restricted experimentally, for example from the absence of
flavour changing neutral currents. For a feasable experimental search,
we must
make some well motivated simplifying asumptions. As shown in
Fig. \fig{unification}, the coupling constants unify within the MSSM. In
the case of local supersymmetry (supergravity), supersymmetry breaking
is communicated from a so-called ``hidden-sector'' to the fields of
the MSSM, the to us ``visible'' sector at or above the unification
scale. If gravity is flavour blind then we expect the supersymmetry
breaking parameters to be universal. This assumption reduces the MSSM
parameter set at the unification scale to six parameters beyond those
of the MSSM:
\beq
{\tilde m}_0,\quad {\tilde m}_{1/2},\quad A,\quad B,\quad \tan\beta,\quad\mu.
\lab{mssmparam}
\eeq
These are respectively, a universal scalar mass, a universal gaugino
mass, universal tri-linear and bi-linear couplings, as well as the
Higgs parameters.

%%%%%%%%%%%%%%%%%%%%%%%%%%%%%%%%%%%%%%%%%%%%%%%%%%
\begin{figure}[t]
\begin{center}
\epsfig{file=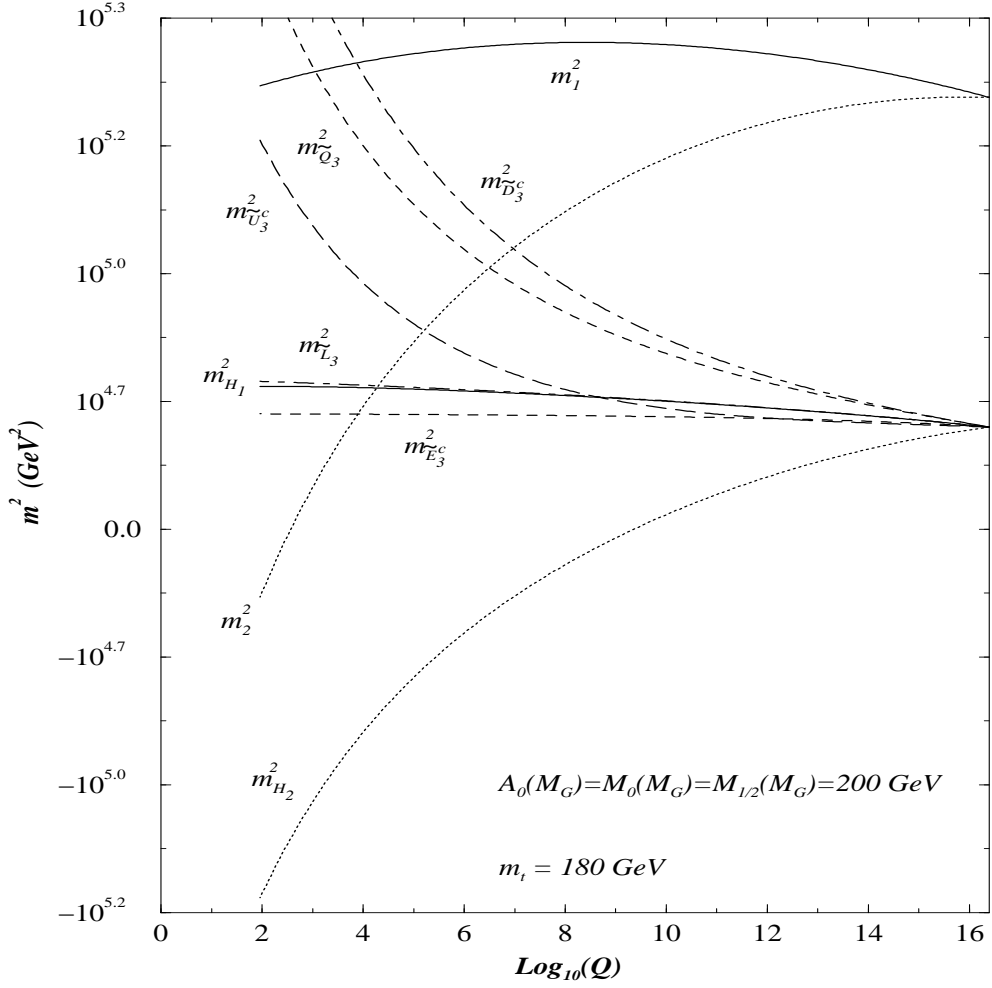,width=13.cm,height=13cm}
\end{center}
\vspace*{-5mm}
\caption[dummy]{{\it Renormalization group evolution of the supersymmetric masses from 
the unification scale down to the electroweak scale.}}
\labf{masses}
\vspace*{-5mm}
\end{figure}
%%%%%%%%%%%%%%%%%%%%%%%%%%%%%%%%%%%%%%%%%%%%%%%%%% 
Given a set of parameters \eq{mssmparam} at the unification scale, the
low-energy spectrum and the couplings are completely predicted through
the supersymmetric renormalization group equations (RGEs). An example
of the running of the masses is shown in\footnote{I thank Athanasios
Dedes for kindly providing me with this figure.} Fig. \fig{masses}. Here 
${\tilde m}_0={\tilde m}_{1/2}= A=200\gev$ at the unification scale.
The various fields run differently since they have different gauge and
Yukawa couplings. A special case is the field $H_2$, whose mass
squared decreases strongly at low energies and becomes negative. This
is not a tachyon. After a suitable field redefinition, shifting the
Higgs field, we see that it corresponds to the spontaneous breaking of
the electroweak symmetry. We can see why this Higgs field is special
from the renormalization group equations
\barr
\frac{dM_H^2}{d\log Q} &=&\frac{1}{4\pi^2}\left[3\lam_t^2(M_H^2+M_{\tilde Q}^2
+M_t^2)+\ldots\right], \\
\frac{dM_{\tilde Q}^2}{d\log Q} &=&
\frac{1}{4\pi^2}\left[2\lam_t^2(M_H^2+M_{\tilde Q}^2
+M_t^2)-\frac{32}{3}g_3^2m^2_{\tilde g}+\ldots\right], \\
\frac{dM_{{\tilde t}_R}^2}{d\log Q} &=&\frac{1}{4\pi^2}\left[3\lam_t^2
(M_H^2+M_{\tilde Q}^2 +M_t^2)-\frac{32}{3}g_3^2m^2_{\tilde g}+\ldots\right],
\earr
where we have only included the dominant terms. The dots refer to
electroweak corrections which are small. $\lam_t$ is
the top quark Yukawa coupling and $g_3$ is the strong coupling
constant.  The Higgs field does not couple strongly and thus for $H_2$
the top-Yukawa coupling dominates the running. This has an opposite
sign from the gauge coupling contribution and drives $M_{H_2}^2$
negative, as $Q^2$ is decreased. In contrast, the running of the squark
masses is dominated by the strong coupling at low-energy and the
masses thus rise with decreasing $Q^2$. In Fig. \fig{masses} this is
shown for the third generation squarks: $M_{{\tilde U}^c_3}$, $M_{{
\tilde D}^c_3}$, and $M_{{\tilde Q}_3}$

As a result of the RGEs, given a set of parameters at the unification
scale we have dynamically generated electroweak symmetry breaking.  We
thus have a ``prediction'' for the $W$ boson mass as a function of the
parameters in Eq.\eq{mssmparam}. Since the $W$ boson mass is already known
we can turn this around to fix a parameter at the unification
scale. It is common to choose $\mu^2$. Since the initial mass squared
for $H_2$ is ${\tilde m}_0^2+\mu^2$ at the unification scale, the sign
of $\mu$ remains undetermined. This mechanism is called `radiative
breaking of electroweak symmetry' \cite{ibanezross}. This solves a
second aspect of the hierarchy problem. Since the RGEs are run
logarithmically the scale at which $M_{H_2}^2$ becomes negative is
many orders of magnitude below the unification scale. Thus we have
dynamically generated an exponential scale difference. In addition,
supersymmetry eliminates the quadratic divergences to stabalize this
large scale hierarchy.

\subsection{Mixing in Supersymmetry}
\label{sect:mixing}
Below the electroweak breaking scale the gauge symmetry is reduced to
$G=SU(3)_C \times U(1)_{EM}$. Thus the superpartners of the $\mu^-$,
the $SU(2) _L$ doublet and singlet fields, ${\tilde\mu}_L^-$ and
${\tilde\mu}_R^-$, respectively, have identical conserved quantum
numbers and can mix. The scalar muon mass matrix squared in the
$({\tilde\mu}_L,{\tilde\mu}_R)$-basis is given by
\beq
{\cal M}^2({\tilde\mu})=\left(
\begin{array}{cc}
m_\mu^2 + {\tilde m}_{\mu_L}^2 -(\half-s_W^2)\cos(2\beta) M_{Z^0}^2
& -m_\mu(A+\mu\tan\beta) \\
-m_\mu(A+\mu\tan\beta) & m_\mu^2 + {\tilde m}_{\mu_L}^2 
-s_W^2\cos(2\beta) M_{Z^0}^2
\end{array}\right).
\eeq
The mixing is proportional to the SM mass term $m_\mu$ and thus
proportional to the Higgs vacuum expectation value, which vanishes in
the $SU(2)_L\times U(1)_Y$ limit. The new mass eigenstates are denoted
\barr
{\tilde\mu}_1&=& {\tilde\mu}_L\cos\theta_{\tilde\mu} + 
{\tilde\mu}_R\sin\theta_{\tilde\mu}, \\
{\tilde\mu}_2&=& -{\tilde\mu}_L\sin\theta_{\tilde\mu} + 
{\tilde\mu}_R\cos\theta_{\tilde\mu}, 
\lab{smumixing}
\earr
and $\theta_{\tilde\mu}$ is the mixing angle which is large for large
off-diagonal terms, \ie large $A$ or large $\tan\beta$. The mixing
matrix is completely analogous for the other $SU(2)$-isospin$=-\half$
fields: ${\tilde e},\,{\tilde\tau}, \,{\tilde d},\,{\tilde
s},\,{\tilde b}$. For the $SU(2)$-isospin$=+\half$ fields, \eg the
${\tilde t}$, the off-diagonal term is $-m_t(A+\mu\cot\beta)$. Thus
large mixing is obtained for large values of $A$ and {\it small} values of
$\tan\beta$.

The electroweak gauge boson and Higgsino fields can also mix. The mass
eigenstates are denoted neutralinos ${\tilde\chi}^0_{i=1,\dots,4}$
which are admixtures of $({\tilde\gamma},{\tilde Z}^0,{\tilde H}_1
^0,{\tilde H}_2^0)$ and charginos ${\tilde\chi}_{i=1,2}^\pm \sim(
{\tilde W}^\pm,{\tilde H}^\pm)$.

The Tevatron collider at Fermilab has run at $\sqrt{s}=1.8\tev$. As
discussed above, we expect the supersymmetric masses to be $\lsim{\cal
O}(1\tev)$. Why have we seen no effects so far? Is it possible for 
supersymmetry to hide just around the corner?

\section{Magnetic Moments}
The Dirac equation for a $\mu$ in an electro-magnetic field $A$ is
given by
\beq
(\not\!\partial-e\not\!\! A-m_\mu)\psi_\mu=0.
\eeq
In an external magnetic field, the Hamiltonian for a $\mu^-$ is given by
\beq
{\cal H}=\frac{e}{m} {\vec S}(\mu)\cdot {\vec B}_{ext},
\eeq
where $\vec S$ is the three-dimensional spin vector, and ${\vec B}$ is
the magnetic field. The Bohr magneton of an electron is defined as
$\mu_B=\frac{e}{2m_e}$. The magnetic moment of the muon is
$\mu_\mu\equiv g_\mu\mu_B$, replacing the electron mass by the muon
mass in $\mu_B$. The Dirac equation predicts $g_\mu=2$ at
tree-level. At one-loop, the QED correction from the graph shown in
Fig. \fig{g2muon}a is \cite{peskin,weinberg}
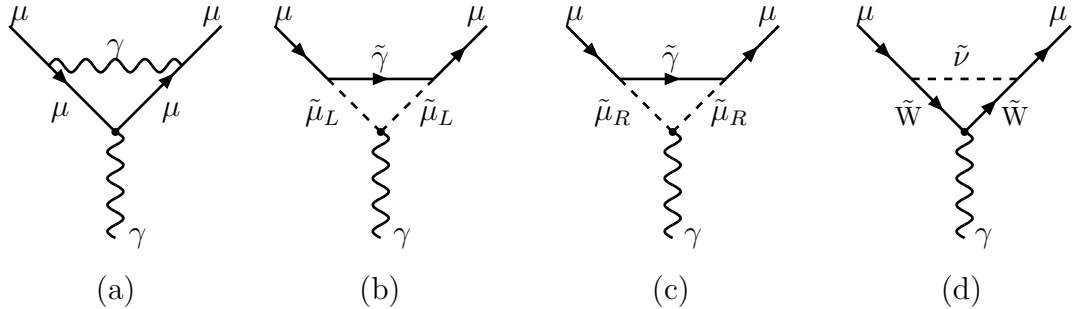
\begin{figure}
\begin{center}
\begin{picture} (300,100)(0,0)
%\thicklines
\SetWidth{1.}
%\SetScale{2.83}
% incoming fermions
\ArrowLine(-70,80)(-30,40)
\ArrowLine(-30,40)(10,80)
\Vertex(-30,40){1.5}
\Photon(-30,40)(-30,0){3}{4}
\Photon(-55,65)(-5,65){2.5}{4.5}
\Text(-70,84)[l]{$\mu$}
\Text(10,84)[r]{$\mu$}
\Text(-25,0)[l]{$\gamma$}
\Text(-30,70)[b]{$\gamma$}
\Text(-47.5,47.5)[r]{$\mu$}
\Text(-12.5,47.5)[l]{$\mu$}

% susy contributions
%left handed smuon
\ArrowLine(30,80)(50,60)
\ArrowLine(50,60)(90,60)
\ArrowLine(90,60)(110,80)
\Vertex(70,40){1.5}
\Photon(70,40)(70,0){3}{4}
\DashLine(50,60)(70,40){3}
\DashLine(70,40)(90,60){3}
\Text(30,84)[l]{$\mu$}
\Text(110,84)[r]{$\mu$}
\Text(75,0)[l]{$\gamma$}
\Text(70,65)[b]{$\tilde\gamma$}
\Text(55,47.5)[r]{${\tilde\mu}_L$}
\Text(85,47.5)[l]{${\tilde\mu}_L$}

% right handed smuon 
 
\ArrowLine(140,80)(160,60)
\ArrowLine(160,60)(200,60)
\ArrowLine(200,60)(220,80)
\Vertex(180,40){1.5}
\Photon(180,40)(180,0){3}{4}
\DashLine(160,60)(180,40){3}
\DashLine(180,40)(200,60){3}
\Text(140,84)[l]{$\mu$}
\Text(220,84)[r]{$\mu$}
\Text(185,0)[l]{$\gamma$}
\Text(180,65)[b]{$\tilde\gamma$}
\Text(165,47.5)[r]{${\tilde\mu}_R$}
\Text(195,47.5)[l]{${\tilde\mu}_R$}

% Chargino contribution

\ArrowLine(250,80)(270,60)
\ArrowLine(310,60)(330,80)
\DashLine(270,60)(310,60){3}
\Vertex(290,40){1.5}
\Photon(290,40)(290,0){3}{4}
\ArrowLine(270,60)(290,40)
\ArrowLine(290,40)(310,60)
\Text(250,84)[l]{$\mu$}
\Text(330,84)[r]{$\mu$}
\Text(295,0)[l]{$\gamma$}
\Text(290,65)[b]{$\tilde \nu$}
\Text(275,47.5)[r]{${\tilde {\mbox{\footnotesize W}}}$}
\Text(305,47.5)[l]{${\tilde {\mbox{\footnotesize W}}}$}
\Text(-30,-20)[c]{(a)}
\Text(70,-20)[c]{(b)}
\Text(180,-20)[c]{(c)}
\Text(290,-20)[c]{(d)}
\end{picture}
\end{center}
\caption{{\it Contributions to the anomalous magnetic moment of the muon.
(a) Standard Model, (b) MSSM for photino and left-handed smuon, (c)
MSSM for photino and right-handed smuon, (d) MSSM for chargino and
sneutrino.}}
\labf{g2muon}
\end{figure}

\beq
\Delta\left(\frac{g_\mu-2}{2}\right)
=\frac{\alpha_{QED}(Q^2=0)}{2\pi}=0.0011614.
\lab{smmuon}
\eeq
This is a one part in $10^3$ correction which is small and independent
of the mass. Experimentally one observes $\Delta(\frac{g_\mu-2}{2})=
1.1\,659\,230 (84)\cdot10^{-3}$ \cite{pdg}, so the one-loop theoretical
result is already close to the experimental value.

The proton is a spin-$\half$ fermion and should also obey the Dirac
equation. We thus expect $(\frac{g_p-2}{2})\lsim 10^{-3}$,
with $\mu_B(p)= \frac{e}{2M_{p}}$, and an analogous result
for the neutron.  However, experimentally we observe \cite{pdg}
\barr
\left(\frac{g-2}{2}\right)_{proton}&=&0.395, \\
\left(\frac{g-2}{2}\right)_{neutron}&=&-1.955.
\lab{magexp}
\earr
These are very large anomalies which can not be accommodated by
radiative corrections. The explanation of this discrepancy is ``new
physics'', namely {\it quarks}. Introducing an $SU(6)$ symmetry of
flavour and spin (thus combining an internal with an external
symmetry!), we require the wave-function of the bound quark state to be
symmetric under the simultaneous interchange of flavour {\it and}
spin. The two up-quarks in the polarized proton should then have a
symmetric spin wave-function, in order to maintain $SU(6)$ symmetry,
\beq
\psi_{proton}^\Uparrow(j=\half,j_z=\half)=
\sqrt{\frac{2}{3}}\;\chi^{uu}_{symm}(1,1)\;\phi^d(\half,-\half)
\;-\;\sqrt{\frac{1}{3}}\; \chi^{uu}_{symm}(1,0)\;\phi^d(\half,\half).
\eeq
The magnetic moment of the proton is then
\beq
\mu_{proton}\propto|\psi_{proton}^\Uparrow|^2 \;=\;
\frac{2}{3}(2\mu_u-\mu_d)+\frac{1}{3}\mu_d\; =\;
\frac{4}{3}\mu_u-\frac{1}{3}\mu_d,
\eeq
Therefore, using $\mu_q=\frac{e_qe}{2m_q}$ we obtain
\beq
g_{proton}=2.79,\quad {\rm and}\quad (\frac{g-2}{2})_{proton}=0.3965,
\eeq
which agrees to within $\frac{\alpha}{2\pi}$ (the SM one-loop 
correction in Eq.\eq{smmuon}) with the experimental number
in Eq.\eq{magexp}. A similar result is obtained for the neutron.

\section{Muon Magnetic Moment in Supersymmetry}
The magnetic moment of the muon is measured and calculated to a 
high accuracy \cite{pdg,muon}
\barr
{\rm EXP:}\quad \left(\frac{g-2}{2}\right)_\mu&=&(11\,659\,230\pm84)10^{-10},
\lab{exp}\\
{\rm TH:} \quad \left(\frac{g-2}{2}\right)_\mu&=&(11\,659\,175\pm16)10^{-10},
\earr
where we have now included the most accurate theoretical calculation.
Thus the allowed range for any supersymmetric contributions is limited
at 90\% C.L. to \cite{carena}
\beq
-90\cdot10^{-10} < \Delta\left(\frac{g-2}{2}\right)_\mu^{susy} 
< 190\cdot 10^{-10}.
\lab{bounds}
\eeq
A new experiment, E821,  at Brookhaven has started and is expected to be
accurate to $4\cdot10^{-10}$, which compared to Eq.\eq{exp} will be a factor 
of 20 better. I would now like to discuss what affect this can have on
searches for supersymmetry.

\subsection{Supersymmetric QED}
In supersymmetric QED there are new contributions at one-loop to the 
anomalous magnetic moment of the muon. The two Feynman diagrams are
shown in Fig. \fig{g2muon}b,\,c. For external muon momenta $p,\;p'$
the contribution of the first graph with virtual ${\tilde\mu}_L$ is given by
\barr
I^\nu&=&\int\frac{d^4k}{(2\pi)^4} (ie\sqrt{2})P_R \frac{1}{\not\! k-M_
{\tilde\gamma}}P_L(ie\sqrt{2}) \frac{i}{(p'-k)^2-m^2_{{\tilde\mu}_L}}
(ie)(p'+p-2k)^\nu \frac{i}{(p-k)^2-m^2_{{\tilde\mu}_L}}\nonumber \\
&=&-2e^3 \int\frac{d^4k}{(2\pi)^4} \frac{P_R(\not\! k+M_{\tilde\gamma})P_L}
{k^2-M^2_{\tilde\gamma}}\frac{(p'+p-2k)^\nu}{[(p'-k)^2-m^2_{{\tilde\mu}_L}]
[(p-k)^2-m^2_{{\tilde\mu}_L}]}
\earr
We now make use of $P_R\not\! kP_L=P_R\not\! k$ and $P=p+p'$.
Furthermore, $P_RM_{\tilde\gamma}P_L=0$. The photino mass term drops
out because of the chirality of the coupling of the fields
${\tilde\gamma}{\tilde\mu}_ L\mu$, which goes as $eP_L$. The smuon
field is an $SU(2)_L$ eigenstate.  Below we will allow for mixing of
the scalar muons. As shown in Eq.\eq{smumixing}, the mass eigenstates
are then no longer eigenstates of the $SU(2)_L$ current and the
coupling is not purely chiral.  This will then result in a term
proportional to $M_{\tilde\gamma}$. For now we assume no mixing of the
smuons, then
\beq
I^\nu=(-2e^3)\int\frac{d^4k}{(2\pi)^4}
\frac{\not \!k (P-2k)^\nu}{[k^2-M^2_{\tilde\gamma}]
[(p'-k)^2-m^2_{{\tilde\mu}_L}]
[(p-k)^2-m^2_{{\tilde\mu}_L}]}.
\lab{susyint}
\eeq
To calculate this integral one follows the same steps as in the QED
calculation which is given in many textbooks, \eg \cite{weinberg,peskin}.
Here we just discuss the general form of the integral. From Lorentz
invariance we expect 
\beq
<\!u(p')|I^\nu|u(p)>=<\!u(p')|
\gamma^\nu C_1+P^\nu C_2+(p-p')^\nu C_3|u(p)>,
\lab{current}
\eeq
where the $C_i$ are Lorentz scalars. Current conservation now implies
the Ward identity $<u(p')|q_\nu I^\nu|u(p)>=0$. From the Dirac
equation we obtain $\not \!p \,u(p)=m\, u(p)$, and ${\ol u}(p')
\not\! p'\,=m\, {\ol u}(p')$. Therefore $<\!u(p')|q_\nu\gamma
^\nu|u(p)\!>= <u(p')|q_\nu P^\nu|u(p)>=0$. Current conservation then
implies that $C_3=0$. This is confirmed by the explicit calculation.
Next we use the Gordon identity
\beq
<\!{\ol u}(p')|\gamma^\nu| u(p)\!>=<\!{\ol u}(p')|\left[ \frac{P^\nu}{2m}
+ \frac{i\sigma^{\nu\rho}q_\rho}{2m}\right]| u(p)\!>,
\eeq
to re-write the term in Eq.\eq{current} proportional to $P^\nu$,
and modifying the coefficient $C_1$ we get
\beq
<\!u(p')|I^\nu(p',p)|u(p)>=<\!u(p')|
\gamma^\mu F_1(q^2) + \frac{i\sigma^{\nu\rho}q_\rho}{2m}
F_2(q^2)|u(p)>.
\eeq
At tree-level the vertex is given by $\gamma^\nu$ and thus $F_1(0)=1$ and
$F_2(0)=0$. At one-loop the corrections to $F_1(0)$ are absorbed in
the renormalization of the electric charge $e^2$. The one-loop
correction to $F_2(0)$ gives the anomalous magnetic moment $(g-2)/2$
\cite{peskin188}.
\beq
\left(\frac{g-2}{2}\right)^{susy}_\mu=F_2(q^2=0)=-\frac{m^2_\mu e^2}{8\pi^2}
\int^1_0dx \frac{x^2(1-x)}{m_\mu^2x^2+(m_{{\tilde\mu}_L}^2-M_{\tilde\gamma}^2 
-m_\mu^2)x+M_{\tilde\gamma}^2}.\lab{susyres}
\eeq
Notice that this is quadratic in $m_\mu^2$. Let us consider the
integral in various limits. In the supersymmetric limit, where
$m_{{\tilde\mu}_L}=m_\mu$ and $M_{\tilde\gamma}=0$ we obtain
\beq
\left(\frac{g-2}{2}\right)_\mu^{{\rm susy}}=-\frac{\alpha_{{\rm em}}}{6\pi}.
\eeq
Recall that in QED we had $+{\alpha_{{\rm em}}}/{2\pi}$ ({\it c.f.}
Eq.\eq{smmuon}). The supersymmetric contribution thus has opposite
sign and is smaller by a factor of 3. In the limit of a light photino,
%%%%%%%%%%%%%%%%%%%%%%%%%%%%%%%%%%%%%%%%%%%%%%%%%%
\begin{figure}[t]
\begin{center}
\vspace*{-0.1cm}
\epsfig{file=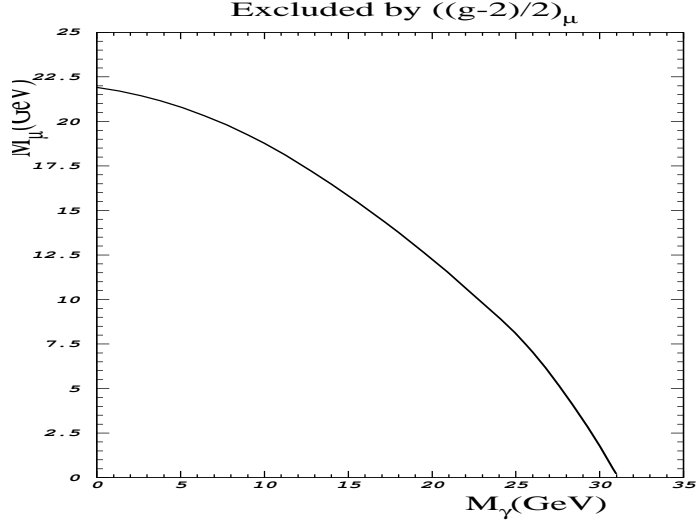,width=9.cm,height=7cm}
\end{center}
\vspace*{-5mm}
\caption[dummy]{{\it Excluded range in the $M_{\tilde\gamma}$
$M_{\tilde\mu}$ mass plane. The region below the curve is excluded.}}
\labf{g2susy}
\vspace*{-5mm}
\end{figure}
%%%%%%%%%%%%%%%%%%%%%%%%%%%%%%%%%%%%%%%%%%%%%%%%%% 
but a heavy scalar muon mass which breaks supersymmetry
($m_{{\tilde\mu}_L}\gg M_{\tilde\gamma},\,m_\mu$), the first
correction is given by
\beq
\left(\frac{g-2}{2}\right)_\mu^{{\rm susy}}=-\frac{\alpha_{{\rm
em}}}{6\pi} \frac{m_\mu^2}{m_{{\tilde\mu}_L}^2}.
\lab{massrat}
\eeq
This is a realistic scenario for supersymmetry breaking. We can now
see how rapidly the supersymmetric contribution decouples in this
case. For $m_{{\tilde\mu}_L}>10 m_\mu/\sqrt{3}$ it is already suppressed by
two orders of magnitude compared to the one-loop SM result in 
Eq.\eq{smmuon}.  We can also see why we have calculated the correction to
the anomalous magnetic moment of the muon and not the electron, even
though the latter is measured to a higher precision \cite{pdg}. The
supersymmetric correction is proportional to the fermion mass squared
and is thus highly suppressed in the case of the electron.
%%%%%%%%%%%%%%%%%%%%%%%%%%%%%%%%%%%%%%%%%%%%%%%%%%
\begin{figure}[t]
\begin{center}
\vspace*{-0.1cm}
\epsfig{file=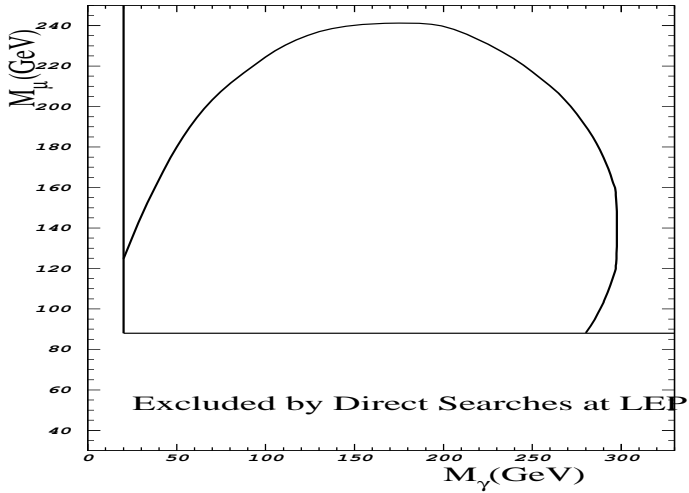,width=9.cm,height=7cm}
\end{center}
\vspace*{-5mm}
\caption[dummy]{{\it Excluded range in the $M_{\tilde\gamma}$
$M_{\tilde\mu}$ mass plane for maximal mixing in the scalar muon
sector and $m_{{\tilde\mu}_1}\ll m_{{\tilde\mu}_2}$. The region
between the curve and the horizontal line is excluded.}}
\labf{g2susymix}
\vspace*{-5mm}
\end{figure}
%%%%%%%%%%%%%%%%%%%%%%%%%%%%%%%%%%%%%%%%%%%%%%%%%% 

In Eq.\eq{bounds} we showed the 90{\%} C.L. limits on the supersymmetric contribution 
to $(g-2)/2$. If we numerically integrate the exact result in
Eq.\eq{susyres} we obtain an excluded region for the only two unknown
parameters $M_{\tilde\gamma}$ and ${\tilde m}_{\mu_ L}$. This is shown
in Fig.\fig{g2susy} using the equations from \cite{grifols}.  These
bounds were first obtained in 1980 \cite{fayet,grifols,history}
and showed that the supersymmetry breaking scale had to be above
$10\gev$. However, as mentioned above, the Brookhaven experiment E821
will have an error reduced by a factor of 20 and thus a mass
sensitivity enhanced roughly by a factor of $\sqrt{20}\approx
4.5$. Thus we expect a maximal mass sensitivity up to
$M_{\tilde\gamma}<135\gev$ and ${\tilde m}_{\mu_L}<100\gev$. These
both exceed the expected LEP2 mass sensitivity.

\subsection{Scalar Muon Mixing}
As discussed in Sect. \ref{sect:mixing}, the scalar muons can
mix. Without mixing the coupling $\gamma{\tilde\mu}_L\mu$ coupled
chiraly as $eP_L$. The mixed mass eigenstate ${\tilde\mu}_1$ couples
as $e(\cos\theta_{\tilde\mu} P_L+\sin \theta_{\tilde\mu} P_R)$. The
first numerator in the integrand of Eq.\eq{susyint} is thus modified
to
\barr
P_L(\not\! k+M_{\tilde\gamma})P_R&\ra&
(c_{\tilde\theta}P_L+s_{\tilde\theta}P_R)
(\not\! k+M_{\tilde\gamma})(c_{\tilde\theta}P_R+s_{\tilde\theta}P_L),
\nonumber \\
&=&\sin\theta_{\tilde\mu}\cos\theta_{\tilde\mu} M_{\tilde\gamma}.
\earr
We have dropped the term proportional to $\not\!k$ since as seen in
the previous calculation it leads to a term proportional to $m_\mu\ll
M_{\tilde\gamma}$. The mass ratio in Eq.\eq{massrat} is then modified to
\beq
\frac{m_{\mu}^2}{{\tilde m}_\mu^2}\ra
-\half\sin\theta_{\tilde\mu}
\cos\theta_{\tilde\mu} 
\frac{m_\mu M_{\tilde\gamma}}{({\tilde m}_\mu^2-M_{\tilde\gamma}^2)^3}
\left[\half({\tilde m}_\mu^4-M_{\tilde\gamma}^4)-{\tilde m}_\mu^2
M_{\tilde\gamma}^2\ln \frac{{\tilde m}_\mu^2}{M_{\tilde\gamma}^2}\right].
\eeq
For moderate to large mixing this is larger than the previous result
by $M_{\tilde\gamma}/m_\mu$. As seen in Eq.\eq{smumixing} the mixing
will be large for large values of $\tan\beta$. We should note that
${\tilde\mu}_1$ and ${\tilde\mu}_2$ contribute with opposite sign to
$I^\nu$. If $m_{{\tilde\mu}_1}=m_{{\tilde\mu}_2}$ they exactly cancel.
This does not occur for mixing. In Fig.\fig{g2susymix}, I show the
excluded region for the case where $m_{{\tilde\mu}_1}=2\cdot m_{
{\tilde\mu}_2}$ and the mixing angle is maximal. The bounds are now an
order of magnitude stronger and substantially above the LEP2 excluded
range. For no mixing we would revert to the previous bounds of Fig. 
\fig{g2susy}. Thus the bounds on $M_{\tilde\gamma}$ and $m_{\tilde
\mu}$ strongly depend on $\tan\beta$. If the experiment E821 performs 
as expected, the mass sensitivity should be pushed up by a factor
$\sqrt{20}\approx4.5$ in both Figs. \fig{g2susy} and \fig{g2susymix}.
Thus masses can be probed, which are well in excess of the LEP2
sensitivity.

%%%%%%%%%%%%%%%%%%%%%%%%%%%%%%%%%%%%%%%%%%%%%%%%%%
\begin{figure}[t]
\begin{center}
\vspace*{-2.5cm}
\epsfig{file=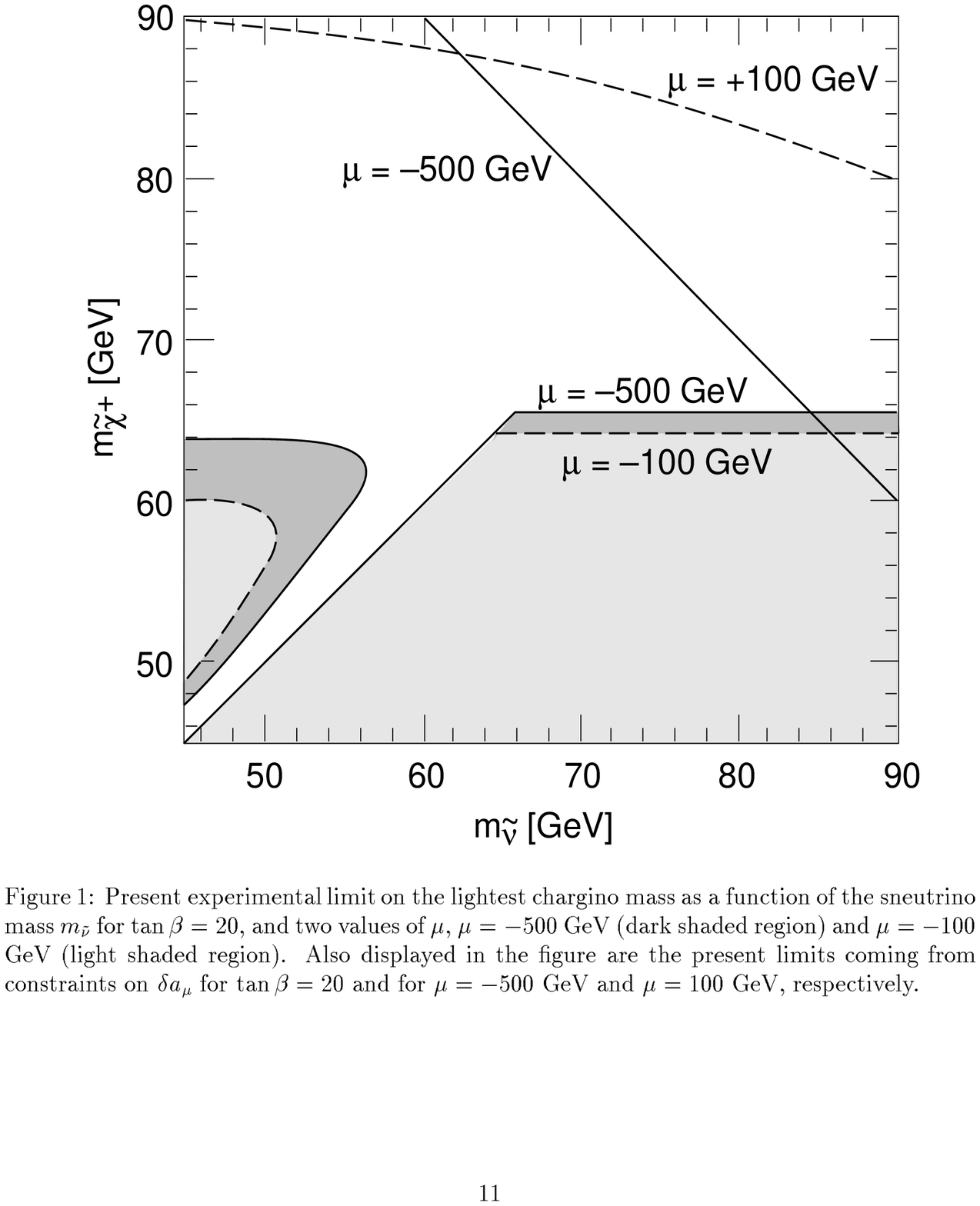,width=15.cm,height=15.cm}
\end{center}
\vspace{-5mm}
\labf{marcela}
\vspace*{-3cm}
\end{figure}
%%%%%%%%%%%%%%%%%%%%%%%%%%%%%%%%%%%%%%%%%%%%%%%%%% 

The full supersymmetric calculation within the MSSM has been presented
in \cite{moroi,carena}. Given a set of MSSM parameters, the full
spectrum is fixed, including that of the smuons. The MSSM calculation
thus includes the mixing for the smuons but also for the neutralinos
and it includes the further diagram due to chargino exchange shown in
Fig.\fig{g2muon}d. The calculation proceeds just as discussed for the
photino case. The chargino diagram has an opposite sign and dominates
for large $\tan\beta$.  The sensitivity in the chargino-sneutrino mass
plane is shown in the Figure on the next page.\footnote{I thank the authors of
\cite{carena} for kindly providing me with this figure.} \cite{carena}. 
The shaded areas were excluded by LEP at the time of the paper in
1996. They have since been extended to $M_{\chi^\pm}>90 \gev, \,
M_{\tilde\nu}\approx 95\gev$ \cite{lepbound}. The {\it current}
qexperimental sensitivity in $(g-2)_\mu /2$ lies below the two curves
labelled by $\mu=+100,\,-500\gev $, and thus below the present LEP2
sensitivity. Note that this plot is for a large value of $\tan\beta=
20$. These curves should be pushed up in sensitivity by more than
$\sqrt{20}$ by the E821 experiment and thus {\it exceed} the maximal
range of LEP2.

\section{Radiative Corrections}
\subsection{The Standard Model}
The electroweak sector in the SM is described by four parameters 
\beq
g_1,\;g_2,\;\lam_\phi,\;\mu^2.
\eeq
Here $g_{1,2}$ are the gauge couplings and $\lam_\phi$ and $\mu^2$ are
the quartic and bi-linear Higgs couplings, respectively. Equivalently,
this set can be described by the parameters
\beq
\alpha,\;M_W,\;M_Z,\;M_H,
\eeq
where now $\alpha$ is the the electromagnetic coupling, $M_W,M_Z$ are
the weak gauge boson masses and $M_H$ is the Higgs mass. The conversion 
to the previous set is given by
\barr
e&=&\sqrt{4\pi\alpha}, \\
\frac{M_W}{M_Z}&=&\cos\theta_W,\quad \Longrightarrow\quad
g_1=\frac{e}{\cos\theta_W},\;g_2=\frac{e}{\sin\theta_W},\\
\mu^2&=&\half M_H^2,\\
v&=&\frac{2M_W}{g_2}=\frac{2M_Z}{\sqrt{g_1^2+g_2^2}},\quad
\Longrightarrow \quad\lam_\phi
=\frac{4\mu^2}{v},\;h_f=\frac{\sqrt{2}\,m_f}{v}.
\earr
All electroweak observables are determined by these 4 parameters plus
the fermion masses.  However, there are more than 25 precision
electroweak measurements. One thus uses 4 measurements plus the
fermion masses to fix the above parameters. Given these 4 parameters
the results for all other 21 measurements can be calculated in the SM
and therefore predicted. The remaining 21 measurements are then a
stringent test of the consistency of the SM.  The precision of
measurements, mainly at LEP, is such that radiative corrections are
tested, \ie in calculating the predictions one must include higher
order corrections.  As an example, consider Fermi's constant as
measured in $\mu$ decay. It is fully determined from the above
parameters
\beq
\left(G_F\right)_\mu =\frac{\pi\alpha}{2(1-\frac{M_W^2}{M_Z^2})M_W^2}\;
\frac{1}{(1-\Delta\alpha)\,(1+\frac{\cos^2\theta_W}{\sin^2\theta_W}
\;\Delta\rho)-(\Delta r)_{\rm rest}}.
\lab{gfermi}
\eeq
Here the first fraction gives the tree-level result. The second
fraction is due to radiative corrections and is equal to one at
tree-level, \ie $\Delta\alpha=\Delta\rho=(\Delta r)_{\rm rest}=0$ at
tree-level. The corrections to $\alpha$, $\Delta\alpha$, are due to the
vacuum polarization of the photon as shown in Figure \fig{2pt}a. The
corrections to the $\rho$ parameter, $\rho\equiv M_W^2/(\cos^2\theta_W
M_Z^2)$, are due to the equivalent vacuum polarisation of the $W$ and $Z$
bosons. The remaining radiative corrections to muon decay, summarized 
in $(\Delta r)_{\rm rest}$ do not factorize into two-point functions. 
An example is shown in Fig. \fig{delrest}a.
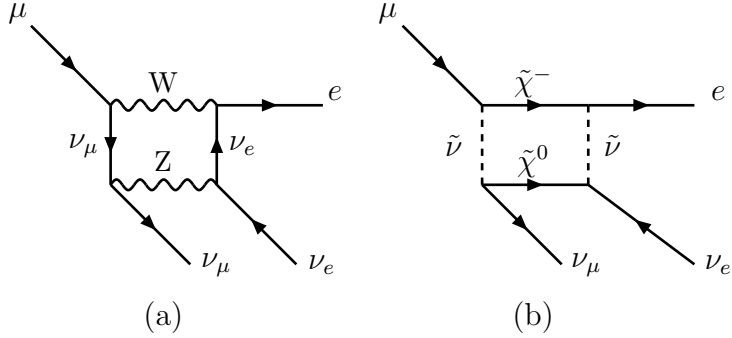
\begin{figure}
\begin{center}
\begin{picture} (300,120)(0,0)
%\thicklines
\SetWidth{1.}
%\SetScale{2.83}
% incoming fermions
\ArrowLine(0,100)(30,70)
\ArrowLine(30,70)(30,40)
\ArrowLine(30,40)(60,10)
\Photon(30,70)(70,70){2}{4.5}
\Photon(30,40)(70,40){2}{4.5}
\ArrowLine(70,70)(110,70)
\ArrowLine(70,40)(70,70)
\ArrowLine(100,10)(70,40)
\Text(-5,105)[c]{$\mu$}
\Text(50,79)[c]{$\mbox{\small W}$}
\Text(50,49)[c]{$\mbox{\small Z}$}
\Text(20,55)[c]{$\nu_\mu$}
\Text(80,55)[c]{$\nu_e$}
\Text(115,75)[c]{$e$}
\Text(110,10)[c]{$\nu_e$}
\Text(70,10)[c]{$\nu_\mu$}

\ArrowLine(140,100)(170,70)
\DashLine(170,70)(170,40){3}
\ArrowLine(170,40)(200,10)
\ArrowLine(170,70)(210,70)
\ArrowLine(170,40)(210,40)
\ArrowLine(210,70)(250,70)
\DashLine(210,40)(210,70){3}
\ArrowLine(250,10)(210,40)
\Text(135,105)[c]{$\mu$}
\Text(190,79)[c]{${\tilde\chi}^-$}
\Text(190,49)[c]{${\tilde\chi}^0$}
\Text(160,55)[c]{${\tilde\nu}$}
\Text(220,55)[c]{${\tilde\nu}$}
\Text(260,75)[c]{$e$}
\Text(210,10)[c]{${\nu}_\mu$}
\Text(260,10)[c]{$\nu_e$}

\Text(50,-10)[c]{(a)}
\Text(190,-10)[c]{(b)}
\end{picture}
\end{center}
\caption{{\it Radiative contributions to muon decay for (a) the SM and
(b) the MSSM. These contributions are included in $\Delta r_{\rm rest}$.}}
\labf{delrest}
\end{figure}

%%%%%%%%%%%%%%%%%%%%%%%%%%%%%%%%%%%%%%%%%%%%%%%%%%
\begin{figure}[t]
\begin{center}
\epsfig{file=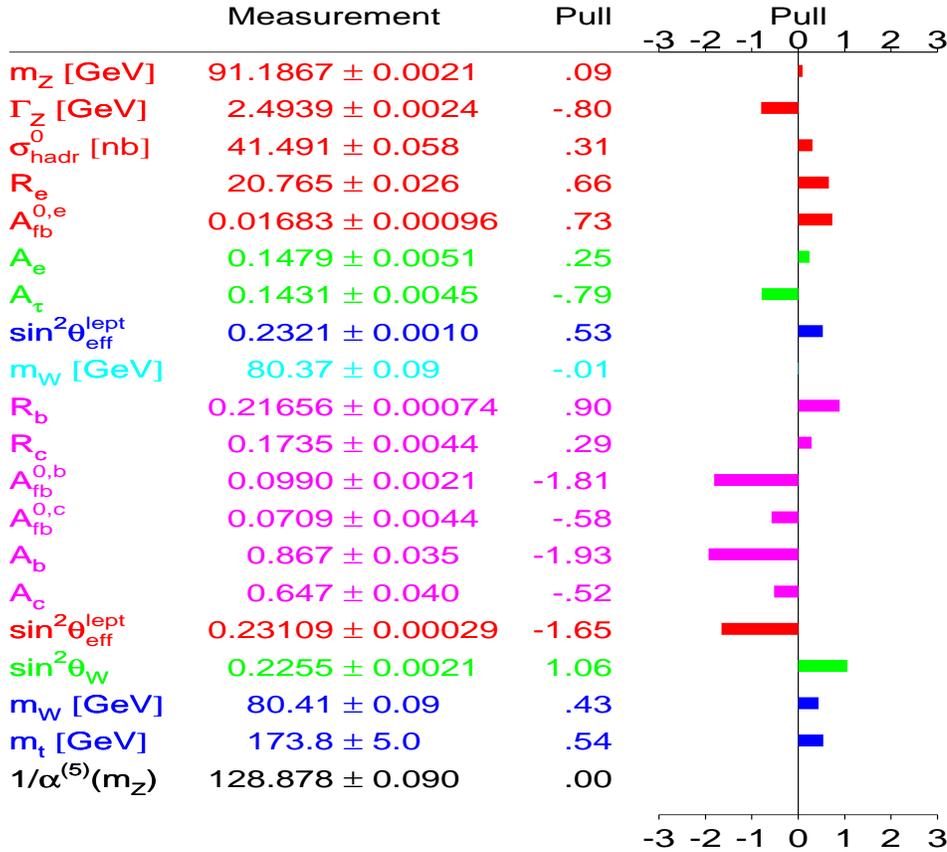,width=13.cm,height=13.5cm}
\end{center}
\vspace*{-10mm}
\caption[dummy]{{\it SM electroweak fit as presented in Vancouver by W. Hollik.}}
\labf{lepfit}
\vspace*{-5mm}
\end{figure}
%%%%%%%%%%%%%%%%%%%%%%%%%%%%%%%%%%%%%%%%%%%%%%%%%% 

All of these corrections are included in the global electroweak fits.
The data as well as the pull of the fit is shown in \footnote{I thank
Wolfgang Hollik for kindly providing me with thsi figure.} Fig. 
\fig{lepfit} \cite{hollik}. All experiments agree with the SM to
better than $2\sigma$.  The $\chi^2/{\rm d.o.f.}=18.4/16$ is excellent
\cite{hollik}. From electroweak precision data there is no hint for
any physics beyond the SM. This is unlike the case of the magnetic
moment of the proton, which gave a strong indication of a new scale of
physics. However, and this is one of the main points of this lecture,
supersymmetry can all the same be hiding just around the corner. As we
saw for $(g-2)_\mu/2$ supersymmetry decouples very rapidly, well
before the upper mass bound of $1\tev$ imposed by the hierarchy
problem. It is therefore no great problem that the electroweak fit is
so good.

%%%%%%%%%%%%%%%%%%%%%%%%%%%%%%%%%%%%%%%%%%%%%%%%%%
\begin{figure}[t]
\begin{center}
\vspace*{1cm}
\epsfig{file=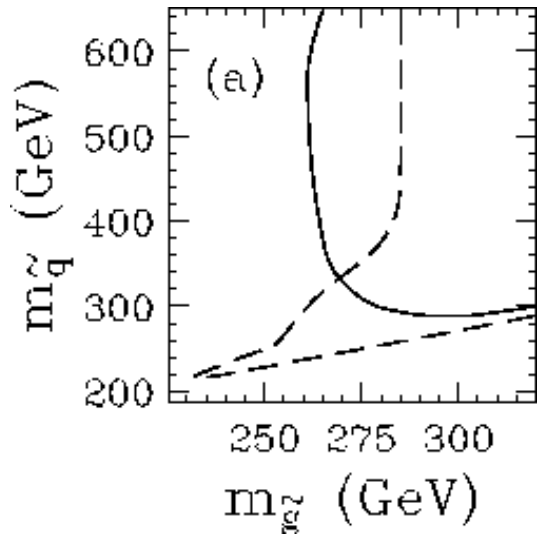,width=7.cm,height=7.cm}
\end{center}
\vspace*{-5mm}
\caption[dummy]{{\it Envelopes of the excluded parameter points at the
$95\%$ C.L. shown as the dashed line and the allowed parameter
points shown as a solid line in the gluino- and squark-mass plane.}}
\labf{damien}
\vspace*{-5mm}
\end{figure}
%%%%%%%%%%%%%%%%%%%%%%%%%%%%%%%%%%%%%%%%%%%%%%%%%% 

\subsection{Radiative Corrections and the MSSM}
Within the MSSM, the parameters receive radiative corrections from the
superpartners as well, just as in the calculation for $(g_\mu-2)/2$.
These modify the predictions for other measurements such as that for
$(G_F)_\mu$ of Eq.\eq{gfermi}. The tree-level result is unmodified,
but there are now additional contributions to (i) $\Delta\alpha$ and
(ii) $\Delta\rho$ through supersymmetric particles in the vacuum
polarization loop. There are also new contributions to $(\Delta
r)_{\rm rest}$ as shown for example in Fig. \fig{delrest}b.

In the minimal supergravity model with radiative breaking we have the 
parameter set 
\beq
\left[\left( \alpha,\,G_F,\,M_Z,\,M_H, m_{top}\right)_{SM};
\left(m_0,\,m_{1/2},\,\tan\beta,\, A_t,\, sgn(\mu)\right)_{MSSM}\right].
\eeq
We thus have 9 parameters plus one sign with again more than 25
observables. This is a non-trivial check of the MSSM, a hurdle which
technicolour theories for example have great difficulty in passing.
The philosophy of the authors in \cite{hollik} is to let the
supersymmetry parameters float freely and thus obtain a preferred
value from an optimal fit. The resulting global fit to the data gives
a $\chi^2/d.o.f=17.3/13$ \cite{hollik}, which is slightly worse than
in the SM case but still an excellent fit. We call the resulting value
$\chi^2_{min}$. As we showed in the case of the anomalous magnetic
moment of the muon, the supersymmetric contributions decouple very
quickly. It is thus no great surprise, that the data can be fit with
heavy supersymmetric masses.

A different philosophy is taken by J. Erler and D. Pierce
\cite{pierce}, which is the same philosophy we used above when
discussing the supersymmetric contributions to $(g_\mu-2)/2$. If we
lower the supersymmetric masses the supersymmetric corrections become
larger.  This is clear in our result for $((g_\mu-2)/2)^{susy}$,
Eq.\eq{massrat}. Since the SM fit is so good this implies that the
overall $\chi^2$-fit becomes worse as we lower the masses. In practice
the authors scan the MSSM parameter space. At each point in the scan,
the constraints from electroweak symmetry breaking, Yukawa
perturbativity and direct searches are applied. Points failing these
checks are disregarded.  For the remaining points, the $\chi^2$ is
computed. If $\Delta\chi^2 =\chi^2-\chi^2_{min}>3.84$, this point is
excluded at 95\% C.L.. The remaining points are allowed. In
Fig. \fig{damien} we show the envelopes of the allowed and excluded
points in the squark gluino mass-plane. We can read off a lower-bound
on the gluino mass of $260\gev$ and of the squark mass of
$285\gev$. The lower bound on the left-handed scalar muon is
$105\gev$. A full set of bounds is given in a Table in Ref.\cite{pierce}.

\section{Supersymmetry Breaking}
\label{sect:susybr}
There is at present no satisfactory model of supersymmetry breaking.
Here I only want to discuss the effect of spontaneously breaking
global supersymmetry on the super trace formula \eq{supertr}. We shall
follow the discussion in Ref.\cite{ferrara}. We thus focus on the
bi-linear couplings of all fields to scalar fields.

\begin{figure}
\begin{center}
\begin{picture} (300,100)(0,0)
%\thicklines
\SetWidth{1.}
%\SetScale{2.83}
% incoming fermions
\Photon(-20,80)(20,50){2}{5}
\Photon(-20,20)(20,50){2}{5}
\Vertex(20,50){2.}
\DashLine(20,50)(60,80){4}
\DashLine(20,50)(60,20){4}
\Text(-28,85)[c]{$A_\mu^a$}
\Text(-28,15)[c]{$A_\mu^b$}
\Text(65,85)[c]{$\phi_i$}
\Text(65,15)[c]{$\phi_j$}

\ArrowLine(100,80)(140,50)
\ArrowLine(100,20)(140,50)
\DashLine(140,50)(180,50){4}
\Text(94,85)[c]{$\psi_i$}
\Text(94,15)[c]{$\psi_j$}
\Text(185,55)[c]{$\phi_k$}

\DashLine(220,80)(260,50){4}
\DashLine(220,20)(260,50){4}
\Vertex(260,50){2.}
\DashLine(260,50)(300,80){4}
\DashLine(260,50)(300,20){4}
\Text(215,85)[c]{$\phi_i$}
\Text(215,15)[c]{$\phi_i$}
\Text(305,85)[c]{$\phi_j$}
\Text(305,15)[c]{$\phi_j$}

\Text(21,-5)[c]{$(a)$}
\Text(145,-5)[c]{$(b)$}
\Text(265,-5)[c]{$(c)$}

\end{picture}
\end{center}
\caption{{\it Supersymmetric couplings contributing to the mass terms 
for (a) the gauge bosons, (b) chiral fermions, and (c) scalar
fermions.}}
\labf{biscalars}
\end{figure}
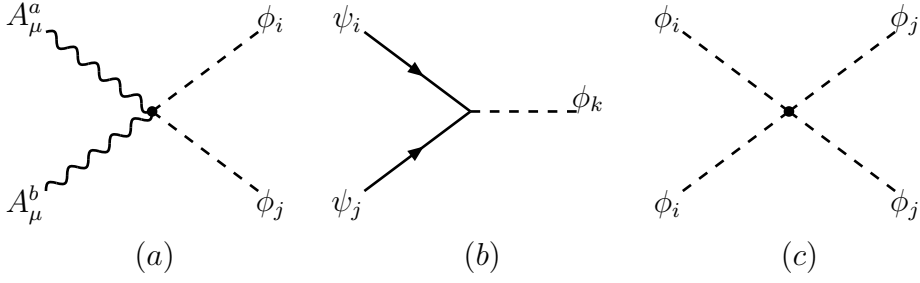

\begin{itemize}
\item {\it Vector Masses:} The vector bosons can obtain masses from
the kinetic term for the scalar fields
\beq
({\cal D}_\mu\phi)^\dagger({\cal D}^\mu\phi),
\eeq
where ${\cal D}_\mu=\partial_\mu +igT_aA^a_\mu$,
is the covariant derivative, $T_a$ is the group
generator and $A^a_\mu$ is the gauge field. The corresponding Feynman
diagram is shown in Fig. \fig{biscalars}a. The coupling is given by $g^2
(T_aT_b)_{ij}A^a_\mu A^b_\nu\phi^*_ i\phi_j$. To obtain the trace of
the mass term we must calculate
$Tr(T_aT_b)_{ij}=C_2(R_\phi)\delta_{ij}$, where $C_2(R_\phi)$ is the
Casimir of the representation of $\phi$ in the gauge group. We thus
obtain
\beq
Tr {\cal M}_V^2=g^2 C_2(R_\phi)\phi^*_i\phi_j. 
\eeq
\item {\it Scalar and Chiral Fermion Masses:} The Yukawa interaction
and the quartic scalar interaction are shown in Fig. \fig{biscalars}b 
and c, respectively. They come from the same superpotential term and are
thus related. For the superpotential term $W_1=LH_1E^c$ which includes
the term giving mass to the electron, we extract the component
interactions via the function ${\overline
W}_1=W_1(\phi)=\phi_L\phi_{H_1}\phi_{E^c}$ which is a function of the
scalar fields, only. The Yukawa coupling is then given by
\beq
\frac{\partial^2{\overline
W}_1}{\partial\phi_L\partial\phi_{E^c}}=\phi_{H_1},\quad {\rm where}
\quad 
m_e\propto <\phi_{H_1}>.
\eeq
The quartic scalar coupling is simply twice the Yukawa coupling.

\item {\it Fermion Masses:} From the previous discussion we obtain for the
chiral fermions 
\beq
m_{ij}=\left|\frac{\partial^2{\overline 
W}}{\partial\phi_i\partial\phi_j} \right|.
\eeq
In the SM the gauge bosons can interact with two fermions, \eg
${\ol\psi}_e\gamma_\mu T^a\psi_e A^\mu_a$. In supersymmetry there is
a corresponding coupling between the superpartner of the gauge boson
(the gauginos, $\lam_{\tilde A}^a$), the SM fermion, $\psi_f$, and the
scalar fermion superpartner, $\phi_{\tilde f}$, namely
$g{\ol\psi}_eP_LT_a\lam_A^a\phi_e$. This gives a mixing between the
gauginos and the chiral fermions. The coupling is proportional to
$g(T^a)_i\phi_i$. In the basis $(\psi,\lam)$ the fermion mass matrix
then has the form
\beq
{\cal M}_F=\left( \begin{array}{cc}
a & b \\
b & 0 \end{array} \right),\quad \Longrightarrow\quad
{\cal M}_F^2 = \left( \begin{array}{cc}
a^2+b^2 & ab \\
ab & 0 \end{array} \right),
\eeq
where $a=\left|\frac{\partial^2{\ol W}}{\partial\phi_i\partial\phi_j}
\right|$ and $b=g T^a_i\phi_i$. For the trace we then
obtain
\beq
Tr{\cal M}_F^2= \left|\frac{\partial^2{\ol W}}{\partial\phi_i\partial\phi_j}
\right| + 2 g^2 C_2(R_\phi)\phi_i^*\phi_j,
\eeq
\item {\it Scalar Masses:} For the scalar masses we obtain
\beq
Tr{\cal M}_S^2=2 \left|\frac{\partial^2{\ol W}}{\partial\phi_i\partial\phi_j}
\right| + g^2 C_2(R_\phi)\phi_i^*\phi_j.
\eeq
The first term is obtained from the superpotential as discussed above.
The second term is obtained from the so-called $D$-term, which also
gives the quartic interactions in the Higgs potential.
\end{itemize}

Next we combine all these terms into the supertrace formula
\barr
STr{\cal M}^2 &=& \sum_{i=s} (-1)^{2S_i}(2S_i+1) Tr {\cal M}_{J_i}^2 
\nonumber \\
&=& 3 Tr{\cal M}_V^2+Tr{\cal M}_S^2-2 Tr{\cal M}_F^2 \nonumber \\
&=& 3\left(g^2C_2(R_\phi)\phi^*_i\phi_i \right) + \left(2 
\left|\frac{\partial^2{\ol W}}{\partial\phi_i\partial\phi_j}
\right|^2+ g^2 C_2(R\phi)\phi^*_i\phi_i\right) \nonumber \\
&&-2\left(
\left|\frac{\partial^2{\ol W}}{\partial\phi_i\partial\phi_j}
\right|^2 + 2 g^2 C_2(R_\phi)\phi^*_i\phi_i\right)\nonumber \\
&=&0.
\earr
We see that all the contributions exactly cancel, {\it independently}
of the value of the scalar fields $\phi_i$. Thus we have $STr{\cal
M}^2=0$, even if some of the scalar fields receive a nonvanishing
vacuum expectation value, $<\phi_i>\not=0$. We therefore have in the
case of spontaneously broken global supersymmetry that
\beq
STr{\cal M}^2=0.
\lab{supertrace0}
\eeq
As discussed in the introduction, this is a catastrophe. For one
chiral multiplet we must have one scalar partner with mass less than the
fermion mass in Eq.\eq{electron}. 

Of course the spectrum is not restricted to just the electron and its
superpartners. In the above derivation we see that the terms deriving
from the superpotential (F-terms) cancel separately from those
proportional to gauge couplings squared (D-terms). In principle a
heavy gaugino could cancel a heavy scalar, letting both be well above
their superpartner masses. However, it has proven impossible to
construct such models \cite{nilles}.

The solution to the above puzzle is given by radiative
corrections. The above formula is only valid at tree-level. It can be
violated by higher order terms. We would expect such corrections to be
small compared to the SM masses, and the supertrace formula to be
violated only weakly. But the required splitting is large, as we saw
for the scalar muon mass. The solution is to embed the supersymmetry
breaking in a so-called ``hidden sector''. This ``sector'' is a set of
fields which do not couple directly to the SM fields.  Here
supersymmetry is broken spontaneously and at leading order the
supertrace vanishes. This breaking is now communicated to the MSSM
superfields via radiative corrections. Thus for these fields the
radiative corrections offer the only and thus dominant contribution to
the breaking mass. They can now have a large splitting from their
SM superpartners such as the muon. 

At present there are two widely considered models. In one case,
supersymmetry breaking is communicated via gauge interactions
\cite{giudice}. In the other case supersymmetry breaking is
communicated via gravity \cite{nilles}. This latter case includes
local supersymmetry, where Eq.\eq{supertrace0} is modified to
\beq
STr{\cal M}^2=2(N-1)m_{3/2}^2,
\lab{supertracegr}
\eeq
where $N$ is the number of chiral superfields and $m_{3/2}$ is
the mass of the gravitino which determines the mass of the
SM superpartners. In both cases we have such a hidden sector,
where the supersymmetry breaking mechanism hides. It is not
clear if this sector will ever be experimentally accessible.
So the disguise of supersymmetry, \ie supersymmetry breaking
giving higher masses as for the smuon, is it self out of sight.

\section{Summary}
Supersymmetry is a well motivated unique extension of the SM. The
hierarchy problem indicates that the supersymmetric masses could
well be of ${\cal O}(1\tev)$.  We discussed the simple example of the
anomalous magnetic moment of the muon and saw that supersymmetry
can indeed be hiding between present experimentally accessible
energy scales and ${\cal O}(1\tev)$ without disrupting precision
measurements, by introducing supersymmetry breaking masses. 
This is the first {\it disguise} of supersymmetry. We then discussed
how the breaking mechanism itself is hidden in order to obtain
a realistic spectrum.

\end{document}